\documentclass[twocolumn,showpacs,preprintnumbers,amsmath,amssymb,superscriptaddress]{revtex4}
\usepackage{graphicx}
\usepackage{ulem}
\usepackage{color}
\usepackage{amssymb}
\input{colordvi.tex}
\begin{document}

\title{Probing small-scale cosmological fluctuations with the 21 cm forest:\\
effects of neutrino mass, running spectral index and warm dark matter}

\author{Hayato Shimabukuro}
\affiliation{%
Department of Physics, Graduate School of Science, 
Nagoya University, Aichi, 464-8602, Japan
}
\email{bukuro@nagoya-u.jp}
\author{Kiyotomo Ichiki}
\affiliation{%
Kobayashi-Maskawa Institute for the Origin of Particles and the
Universe, Nagoya University, Aichi, 464-8602, Japan
}
\author{Susumu Inoue}
\affiliation{%
Max-Planck-Institut f\"ur Physik, 80805 M\"unchen, Germany}
\affiliation{%
Institute for Cosmic Ray Research, University of Tokyo, Kashiwa, Chiba, 277-8582, Japan}
\author{Shuichiro Yokoyama}
{\affiliation{%
Institute for Cosmic Ray Research, University of Tokyo, Kashiwa, Chiba, 277-8582, Japan}

\date{\today}
\begin{abstract}
Although the cosmological paradigm based on cold dark matter and adiabatic, nearly scale-invariant primordial fluctuations
is consistent with a wide variety of existing observations, it has yet to be sufficiently tested on scales
smaller than those of massive galaxies, and various alternatives have been proposed that differ significantly
in the consequent small-scale power spectrum (SSPS) of large-scale structure.
Here we show that a powerful probe of the SSPS at $k\gtrsim 10$ Mpc$^{-1}$
can be provided by the 21 cm forest,
that is, systems of narrow absorption lines due to intervening, cold neutral hydrogen
in the spectra of high-redshift background radio sources in the cosmic reionization epoch.
Such features are expected to be caused predominantly by collapsed gas in starless minihalos,
whose mass function can be very sensitive to the SSPS.
As specific examples, we consider the effects of neutrino mass, running spectral index (RSI)
and warm dark matter (WDM) on the SSPS,
and evaluate the expected distribution in optical depth of 21 cm absorbers out to different redshifts.
Within the current constraints on quantities such as the sum of neutrino masses $\sum m_\nu$,
running of the primordial spectral index $d n_s/d \ln k$ and WDM particle mass $m_{\rm WDM}$,
the statistics of the 21 cm forest manifest observationally significant differences that become larger at higher redshifts.
In particular, it may be possible to probe the range of $m_{\rm WDM} \gtrsim 10$ keV that may otherwise be difficult to access.
Future observations of the 21 cm forest by the Square Kilometer Array may offer a unique and valuable probe of the SSPS,
as long as radio sources such as quasars or Population III gamma-ray bursts with sufficient brightness and number
exist at redshifts of $z \gtrsim$ 10 - 20, and the astrophysical effects of reionization and heating can be discriminated.
\end{abstract}

\preprint{ICRR-Report-670-2013-19}

\maketitle


\section{Introduction}

Over the last decades, cosmological observations have provided us with a wealth of information
about the structure and evolution of the universe.
Dedicated studies of anisotropies in the cosmic microwave background (CMB)
by the {\it COBE}, {\it WMAP} and {\it Planck} satellites as well as ground-based telescopes
have yielded increasingly precise information
on the spectrum of cosmic density fluctuations on the largest scales
\cite{Smoot:1992td, Ade:2013zuv}. 
Extensive surveys of galaxies and clusters of galaxies and their gravitational lensing effects
have clarified the power-spectrum of large-scale structure (LSS) on somewhat smaller scales \cite{Anderson:2012sa}.
Finally, investigations of inhomogeneities in intergalactic hydrogen
through the Lyman alpha forest have led to valuable constraints
on the distribution of cosmic structure on the smallest scales to date \cite{Busca:2012bu}.

Most current observations can be accommodated 
consistently by the concordance $\Lambda$CDM cosmological model,
based on cold dark matter (CDM), a cosmological constant,
and a power-law spectrum of adiabatic primordial density fluctuations that is nearly scale-invariant \cite{Ostriker:1995su}.
However, the $\Lambda$CDM model has yet to be sufficiently tested against observations
on scales much smaller than those corresponding to massive galaxies.
The small-scale power spectrum (SSPS), that is, the power spectrum of large-scale structure
on such scales, is of great interest from different perspectives,
not only for cosmology but also for fundamental physics.

On the premise that the concordance $\Lambda$CDM cosmology is basically valid,
one aspect of the SSPS that has received great attention is its role in constraining the mass of neutrinos
\cite{Lesgourgues:2006nd}. 
The existence of finite rest mass of neutrinos has been demonstrated by experiments probing
solar and atmospheric neutrino oscillations, which are sensitive to the relative mass differences between different neutrino families.
An important cosmological effect caused by light, massive neutrinos is the suppression of the matter power spectrum
on scales smaller than the free streaming scale,
which becomes stronger when the total mass of neutrinos is larger. 
In turn, various cosmological observations including the SSPS can provide valuable upper limits on the total mass of neutrinos.
Current such limits on the sum of neutrino masses of all families is conservatively of order $\sum m_\nu \lesssim 1~{\rm eV}$.
It is of great interest whether further observations can improve on them or even provide measurements of the neutrino mass.

Going somewhat beyond the simplest $\Lambda$CDM model,
an interesting possibility is that the spectrum of primordial fluctuations is not a pure power-law
but has a running spectral index (RSI), that is, a spectral index $n_s$ that is scale-dependent. 
Standard, single-field, slow-roll models of cosmological inflation in the early universe
predict a nearly scale-invariant spectrum with $n_s \simeq 1$,
where $n_s -1$ and $d n_s/d \ln k$ are expected to be first and second order respectively in the small, slow-roll parameters.
However, in some inflation models, relatively large RSI can be realized \cite{Adshead:2010mc}. 
Current cosmological observations indicate $n_s - 1 = O(0.01)$, consistent with standard, slow-roll inflation models.
More extensive observations over a wider range of scales including the SSPS should lead to more precise constraints
and help to discriminate among inflation models that are indistinguishable in terms of $n_s$ alone.

Finally, as a more drastic alternative to CDM, warm dark matter (WDM) with particles masses in the keV range
has been proposed on various grounds \cite{Markovic:2013iza}.
The cosmological effect of WDM is characterized by its free streaming scale that depends on its mass $m_{\rm WDM}$.
Whereas above this scale, it behaves in a way similar to CDM and is indistinguishable from it,
below this scale, it dramatically suppresses the power spectrum, resulting in much fewer dark matter halos on small scales
compared to CDM.
One specific particle physics candidate for WDM is sterile neutrinos,
which are currently constrained to have a mass somewhere in the range $m_{\rm WDM} \sim 1-50$ keV
\cite{Boyarsky:2012rt}. 
Furthermore, WDM has also been motivated from an astrophysical viewpoint.
Current observations indicate that the abundance of satellite galaxies around the Milky Way and in the Local Group
is much lower than compared to the abundance of CDM subhalos on corresponding scales,
the so-called ``missing satellites problem'' \cite{Klypin:1999uc}. 
Although the resolution may lie in astrophysical feedback effects that preferentially quench star formation
in smaller systems, WDM can provide an intriguing alternative explanation
by attributing the lack of satellite galaxies to the absence of the relevant dark halos in such cosmologies.
Such an interpretation may favor a WDM mass in the range of a few keV
 \cite{Polisensky:2010rw} 
(see however \cite{Shao:2012cg}). 
Further observations of the SSPS can offer a critical test of WDM as a viable dark matter candidate.

In order to investigate the SSPS in greater depth and test $\Lambda$CDM and its alternatives,
the most direct approach would be to observationally probe dark matter halos on scales
much smaller than those of galaxies in the present universe.
However, this is made difficult by the fact that the bulk of the intergalactic medium (IGM)
has been fully ionized after completion of cosmic reionization at $z \sim 6$.
In such circumstances, the gas in sufficiently small halos, in particular
``minihalos'' with masses $M \lesssim 10^8 M_\odot$ and virial temperature $T_{vir} \lesssim 10^4$ K,
is expected to have been substantially photoevaporated \cite{Shapiro:2003gxa}.
Thus the only way to probe such halos in the present day
may be via gravitational lensing effects \cite{Dalal:2001fq} 
or possibly DM annihilation gamma rays \cite{Kuhlen:2008aw}, which are quite challenging prospects.

On the other hand, an alternative avenue may open up by focusing on redshifts $z > 6$,
before cosmic reionization and minihalo photoevaporation have proceeded significantly.
At such epochs, the cold, neutral gas associated with collapsed systems may be observable
as a series of redshifted absorption features due to the 21 cm transition
in the continuum spectrum of luminous background radio sources
such as radio quasars or possibly gamma-ray bursts (GRBs).
Dubbed the ``21 cm forest'' in analogy with the Lyman alpha forest,
previous work has shown that the gas in minihalos can give rise to numerous, narrow absorption features
that may be observable with future facilities such as the Square Kilometer Array (SKA) \footnote{http//www.skatelescope.org},
as long as sufficiently bright radio sources exist at the relevant redshifts
\cite{Furlanetto:2002ng, Furlanetto:2006dt, Carilli:2002ky}.
The mass function of minihalos is dependent on the SSPS on scales $k\gtrsim 10$ Mpc$^{-1}$,
much smaller than the smallest scales that are at present observationally accessible via the Lyman alpha forest.
Therefore, future observations of the 21 cm forest can potentially provide a very sensitive probe of the SSPS,
which in turn can provide valuable constraints or measurements of fundamental physics parameters
such as the neutrino mass, the WDM particle mass or RSI of primordial fluctuations,
the prospects of which are the main topic of this paper.

Note that minihalos themselves are unlikely to harbor appreciable star formation,
as their virial temperatures are below the threshold for efficient gas cooling via atomic transitions.
However, the 21 cm forest signal can also be significantly affected by external astrophysical effects,
such as a background of UV photons or heating of the gas via X-rays or shocks,
which are expected to become progressively more important as cosmic reionization proceeds.
Indeed, the implications of such astrophysical feedback processes have been the main focus
of studies on the 21 cm forest so far
\cite{Furlanetto:2002ng, Furlanetto:2006dt, Carilli:2002ky, Xu:2009dr, Xu:2010us, Meiksin:2011gx, Mack:2011if, Ciardi:2012ik, Ewall-Wice:2013yta}
(see however \cite{Vasiliev:2013cba}).
Since the consequences for the 21 cm forest of the SSPS beyond the standard $\Lambda$CDM cosmology
is being discussed here for the first time, as an initial step, here we choose not to account for the
complicating effects of a UV background or heating caused by astrophysical sources.
Thus we are able to isolate and clarify the effects of modifications to the SSPS itself.
The neglect of feedback effects would be more justifiable at higher redshifts, $z \gtrsim 20$,
where the formation and evolution of stars and other objects are expected to be more limited.

We mention that the global signal of 21 cm emission from minihalos and/or the IGM at high redshifts
(for reviews, see \cite{Furlanetto:2006jb, Morales:2009gs}) 
has been previously discussed as a potentially powerful probe of the SSPS \cite{Loeb:2003ya}. 
However, the major obstacle to such prospects is the huge level of foreground emission,
several orders of magnitude brighter than the expected signal, that must be removed very efficiently
in order to observe such emission \cite{Oh:2003jy}.
In contrast, foregrounds are not a concern for observing the 21 cm forest,
as long as sufficiently bright background radio sources exist.

This paper is organized as follows.
In Section II, we describe our basic assumptions and formulation
regarding the halo gas profile, spin temperature, optical depth and abundance of absorbers.
Section III continues with how we formulate the modifications to the SSPS caused by massive neutrinos, RSI and WDM,
and also presents the corresponding expectations for the 21 cm forest,
in comparison with the standard $\Lambda$CDM case.
We end with discussions on the observability, various caveats, and a summary in Section IV.

For our baseline cosmological model, we adopt $\Lambda$CDM with the
following parameters from the latest Planck data:
$\Omega_{m}=0.3175$, $\Omega_{b}h^{2}=0.12029$, $\Omega_{\Lambda}$=0.68,
$H_{0}=100h [{\rm km/s/Mpc}]$, $h=0.67$, $\sigma_{8}=0.834$}, where
$\Omega_{{\rm m}}, \Omega_{{\rm b}}, \Omega_{\Lambda}$ are the
densities of cold dark matter, baryons, and cosmological constant, respectively,
in units of the critical density, $H_{0}$ is the Hubble constant,
and $\sigma_{8}$ is the variance of mass fluctuations averaged over a
sphere with radius 8$h^{-1}{\rm Mpc}$ \cite{Ade:2013zuv}.


\section{Basic formulation}

For the most part, our basic formulation follows that given by Furlanetto \cite{Furlanetto:2002ng}, 
with due modifications for our purposes.

\subsection{Halo gas profile}
We start with the description of the gas density profile in dark matter halos.
We assume that the dark matter potential is described by the
Navarro, Frenk $\&$ White (NFW) profile \cite{Navarro:1996gj, Abel:2000tu},
characterized by the concentration parameter ${\it y}=r_{{\rm vir}}/r_{{\rm s}}$, where $r_{s}$ is the scaling radius, and the virial radius
$r_{{\rm vir}}$ is given by \cite{Barkana:2000fd}
\begin{multline}
  r_{{\rm vir}}=0.784\bigg(\frac{M}{10^{8}h^{-1}M_{\odot}}\bigg)^{1/3}\bigg[\frac{\Omega_{m}}{\Omega_{m}^{z}}
  \frac{\Delta_{c}} {18\pi^{2}}\bigg]^{-1/3}\\
  \times \bigg(\frac{1+{\it z}}{10}\bigg)^{-1}h^{-1}[{\rm kpc}],
\label{eq:virial_radius}
\end{multline}
where $\Delta_{c}=18\pi ^{2}+82d-39d^{2}$ is the overdensity of halos collapsing at redshift ${\it z}$,
with $d=\Omega_{m}^{z}-1$ and
$\Omega_{m}^{z}=\Omega_{m}(1+z)^{3}/(\Omega_{m}(1+z)^{3}+\Omega_{\Lambda})$
\cite{Bryan:1997dn}. 
Here we follow the N-body simulation results of Gao et al. \cite{Gao:2005hn}
for halos at high-redshift and assume that $y$ is inversely proportional to $(1+z)$.

Within the dark matter halo, the gas is assumed to be isothermal and in hydrostatic equilibrium,
in which case its profile can be derived analytically
\cite{Makino:1997dv,Xu:2010us}.
The gas density profile is given by
\begin{equation}
  \ln \rho_{{\rm g}}(r)=\ln \rho_{{\rm g0}}-\frac{\mu m_{{\rm p}}}{2k_{{\rm B}}T_{{\rm vir}}}[v_{{\rm esc}}^{2}(0)-v_{{\rm esc}}^{2}(r)],
\label{eq:gas_profile1}
\end{equation}\\
where
\begin{multline}
  T_{{\rm vir}} =1.32\times 10^{4}\bigg(\frac{\mu}{0.6}\bigg)\bigg(\frac{M}{10^{8}h^{-1}M_{\odot}}\bigg)^{2/3}\\
  \times \bigg[\frac{\Omega_{m}}{\Omega_{m}^{z}}\frac{\Delta_{c}}{18\pi^{2}}\bigg]^{1/3}\bigg(\frac{1+z}{10}\bigg)[{\rm K}]
\end{multline}
is the virial temperature \cite{Barkana:2000fd},
$\rho_{{\rm g0}}$ is the central gas density, $m_{{\rm p}}$ is the
proton mass, and $\mu=1.22$ is the mean molecular weight of the gas.
The escape velocity $v_{{\rm esc}}(r)$ is described by
\begin{equation}
  v_{{\rm esc}}^{2}(r) = 2\int_{r}^{\infty}\frac{GM(r^{'})}{r^{'2}}dr^{'}=2V_{c}^{2}\frac{F(yx)+yx/(1+yx)}{xF(y)}~,
\label{eq:esc_velocity}
\end{equation}
where $x\equiv r/r_{{\rm vir}}$ and $F(y)=\ln(1+y)-y/(1+y)$, and
$V_{c}$ is the circular velocity given by \cite{Barkana:2000fd}
\begin{multline}
  V_{c}^{2} =  \frac{GM}{r_{{\rm vir}}}=23.4\bigg(\frac{M}{10^{8}h^{-1}M_{\odot}}\bigg)^{1/3}
  \bigg[\frac{\Omega_{m}}{\Omega_{m}^{z}}\frac{\Delta_{c}}{18\pi^{2}}\bigg]^{1/6}\\
  \times \bigg(\frac{1+z}{10}\bigg)^{1/2}[{\rm km/s}].
\label{eq:cir_velocity}
\end{multline}

The escape velocity reaches its maximum of $v_{{\rm
esc}}^{2}(0)=2V_{c}^{2}y/F(y)$ at the center of the halo.
The central density $\rho_{g0}$ is normalized by the cosmic value of $\Omega_{b}/\Omega_{m}$ and given by
\begin{equation}
  \rho_{g0}(z) =
  \frac{(\Delta_c/3)y^{3}e^{A}}{\int_{0}^{y}(1+t)^{A/t}t^{2}dt} \left(\frac{\Omega_b}{\Omega_m}\right)\bar{\rho}_{m}(z)~,
\end{equation}
where $A=3y/F(y)$ and $\bar{\rho}_{m}(z)$ is the mean total matter
density at redshift $z$.

In this work, we choose to consider only the gas within $r_{\rm vir}$ for simplicity,
even though the additional gas outside $r_{\rm vir}$ and accreting onto the halo can enhance the absorption signal
\cite{Furlanetto:2002ng, Xu:2010us}.
We also note that in treating nonstandard cosmological effects caused by massive neutrinos, RSI or WDM (Section III),
we focus on the consequent modifications to the halo mass function, and do not account for modifications
to the halo profile itself (see Section IV for further discussion.)


\subsection{Spin temperature}
The optical depth to 21~cm absorption is determined by the HI column density and the excitation
state of the hyperfine transition of the HI atom, characterized by the spin temperature $T_s$. 
Most generally, $T_s$ is determined by the following equation
describing the balance between absorption/emission of CMB photons, collisions with other particles and scattering with UV photons
\cite{Furlanetto:2006jb}:
\begin{equation}
  T_{{\rm S}}^{-1}=\frac{T_{\gamma}^{-1}+x_{c}T_{{\rm K}}^{-1}+x_{\alpha}T_{{\rm C}}^{-1}}{1+x_{c}+x_{\alpha}}.
\label{eq:spin1}
\end{equation}
Here, $T_{\gamma}=2.73(1+{\it z})$ is the CMB temperature at
redshift {\it z}, $T_{{\rm K}}$ is the gas kinetic temperature,
$T_{{\rm C}}$ is the effective color temperature of the UV radiation field,
and $x_{{\rm c}}$ and $x_{{\alpha}}$ are the coupling coefficients for collisions with particles and UV photons, respectively.

As discussed in Section I, in this work we choose not to account for any UV radiation field
in order to understand better how the 21cm forest is affected by modification to the SSPS itself.
Hence the relevant equation simplifies 
\begin{equation}
  T_{{\rm S}}^{-1}=\frac{T_{\gamma}^{-1}+x_{c}T_{{\rm K}}^{-1}}{1+x_{c}}.
\label{eq:spin2}
\end{equation}

We also set $T_{{\rm K}}=T_{{\rm vir}}$, which should be a good approximation
for minihalos where gas cooling is inefficient.
In principle, H$_2$ molecular cooling can reduce the gas temperature slightly below $T_{{\rm K}}$
\cite{Abel:2000tu} and increase the absorption optical depth, but this effect can be mitigated
by H$_2$ destruction by a Lyman-Werner background \cite{Ahn:2008gh}.

The main contributions to collisional coupling are H-H and
H-e$^{-}$ collisions, and $x_{{\rm c}}$ is written as  
\begin{equation}
  x_{{\rm c}}=x^{{\rm HH}}_{{\rm c}}+x^{{\rm He}}_{{\rm c}}
  =\frac{n_{{\rm H{\sc I}}}\kappa_{10}^{{\rm HH}}}{A_{10}}\frac{T_{*}}{T_{\gamma}}
  +\frac{n_{{\rm e}}\kappa_{10}^{{\rm eH}}}{A_{10}}\frac{T_{*}}{T_{\gamma}},
\end{equation}
where $n_{{\rm H{\sc I}}}$ is the number density of H{\sc I} atoms,
$\kappa_{10}^{{\rm HH}}$ and $\kappa_{10}^{{\rm He}}$ are the
de-excitation rates of H-H and H-e collisions, respectively, $A_{10}$
=$2.85\times 10^{-15}$ ${\rm s}^{-1}$ is the Einstein coefficient for
the spontaneous decay of the 21~cm transition, and $T_{*}=0.0682$ K is
the equivalent temperature corresponding to the difference of the energy
levels in the 21~cm transition.
Since we do not account for the effects of cosmic reionization,
the fraction of free electrons is small so that we can neglect the contribution from H-e collisions.
The de-excitation rate $\kappa^{{\rm HH}}$ for $T_{{\rm K}}\le300$ K is
given by Zygelman \cite{2005ApJ...622.1356Z}. For the temperature
range of $T_{{\rm K}} \ge 300$ K, an approximate formula has been
adopted in the literature \cite{Furlanetto:2006jb}.

\begin{figure}
   \centering \includegraphics[width=16cm]{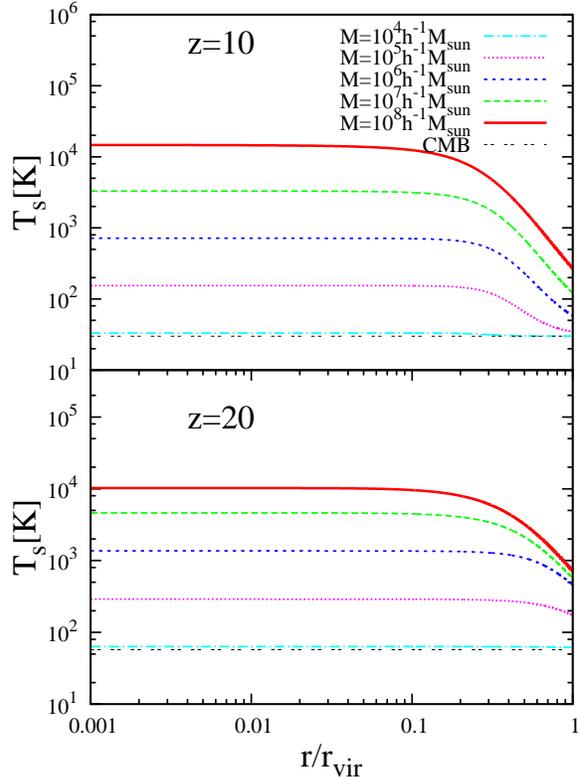}
   \caption{Profiles of spin temperature for minihalos at $z$=10(top), 20(bottom) and different masses as indicated in the legend,
   versus radius normalized by the virial radius. The CMB temperature is shown by the horizontal line.}
\label{fig:spin1}
\end{figure}

The profiles of spin temperature for different minihalo masses at $z=$10 and 20 are shown in Fig.\ref{fig:spin1}.
The spin temperature is equivalent to the virial temperature in the inner regions of
minihalos ($r \ll r_{\rm vir}$), and decreases asymptotically to the CMB
temperature toward the virial radius. 
This is due to the lower number density of the H{\sc I} gas in the outer regions
that makes collisional coupling less effective.

\subsection{Optical depth}
\label{subsec:optical depth}
The optical depth to 21 cm absorption of a halo of mass $M$ at impact parameter
$\alpha$ at frequency $\nu$ can be written as an integral along the line of sight \cite{Furlanetto:2002ng}:
\begin{multline}
  \tau(\nu,M,\alpha)=\frac{3h_{{\rm p}}c^{3}A_{10}}{32\pi k_{{\rm B}}\nu_{21}^{2}}
  \int_{-R_{{\rm max}}(\alpha)}^{R_{{\rm max}}(\alpha)}dR\frac{n_{{\rm H{\sc I}}}(r)}{T_{{\rm S}}(r)\sqrt{\pi}b}\\
  \times \exp\bigg(-\frac{v^{2}(\nu)}{b^{2}}\bigg),
\label{eq:optical1}
\end{multline}
where the velocity dispersion $b=\sqrt{2k_{B}T_{{\rm vir}}/m_p}$.

\begin{figure}
    \centering \includegraphics[width=16cm]{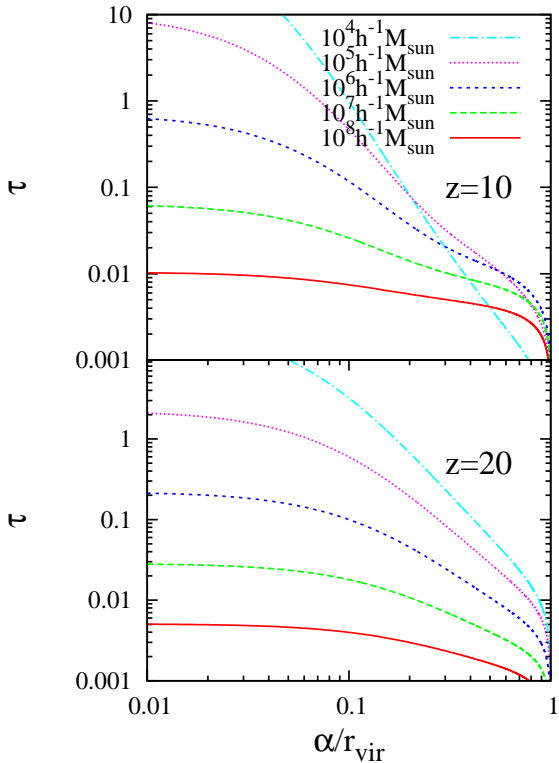}
    \caption{Optical depths to 21 cm absorption for minihalos at $z$=10(top), 20(bottom) and different masses
    as indicated in the legend, as functions of the impact parameter normalized by the virial radius.
   }
\label{fig:tau}
\end{figure}

In Fig.\ref{fig:tau}, the optical depths to 21 cm absorption for minihalos of different masses at $z=$10 and 20
are shown as functions of the impact parameter.
Smaller impact parameters result in larger optical depths 
by virtue of the larger column density despite the higher spin temperature.
We also note that minihalos with smaller masses generally give larger optical depths
when compared at the same impact parameter, which can be understood as follows.
The halo mass dependence of the optical depth in Eq. (\ref{eq:optical1})
comes from $n_{\rm H{\sc I}}$, $T_S$ and $R_{\max}$.
Roughly speaking, we can estimate $n_{\rm H{\sc I}} \sim M / r_{\rm vir}^2$ and $T_S \sim T_{\rm vir}$.
As $r_{\rm vir} \propto M^{1/3}$ and $T_{\rm vir}  \propto M^{2/3}$,
we find $n_{\rm H{\sc I}} \propto M^{1/3}$ and $T_S \propto M^{2/3}$,
pointing to larger optical depths for less massive halos.
The actual dependence may be somewhat more complicated due to the non-trivial density profile.
The width of the absorption feature is determined by $T_S \sim T_{\rm vir}$ and expected to be
of order a few kHz in observer frequency \cite{Furlanetto:2002ng}.


\subsection{Abundance of absorbers}
In order to evaluate the expected abundance of absorption features per redshift interval along an average line of sight,
we introduce a cumulative function
\begin{equation}
  \frac{dN(>\tau)}{dz}=\frac{dr}{dz}\int_{M_{{\rm min}}}^{M_{{\rm max}}}dM\frac{dN}{dM}\pi r^{2}_{\tau}(M,\tau), \label{eq:abundance1}
\end{equation}
where $dr/dz$ is the comoving line element, $r_{\tau}(M,\tau)$ is the maximum impact parameter in comoving units
that gives optical depths greater than $\tau$,
and $dN/dM$ is the halo mass function
representing the comoving number density of collapsed dark matter halos with mass between $M$ and $M+dM$,
here given by the Press-Schechter formalism \cite{Press:1973iz}
(see Section IV for more discussion on the halo mass function.)
The maximum mass $M_{{\rm max}}$ for minihalos is taken to correspond to $T_{{\rm vir}}=10^{4}$ K,
below which gas cooling via atomic transitions and consequent star formation is expected to be inefficient.
The minimum mass $M_{{\rm min}}$ is assumed to be the Jeans mass determined by the IGM temperature \cite{Madau:2003ee},
\begin{equation}
  M_{{\rm J}} =\frac{4\pi \bar{\rho}}{3}\bigg(\frac{5\pi k_{{\rm B}}T_{\rm IGM}}{3G\bar{\rho}m_{{\rm p}}\mu}\bigg)^{3/2}
  \simeq 3.58 \times 10^{5}h^{-1}M_{\odot}\bigg(\frac{T_{\rm IGM}/{\rm K}}{1+z}\bigg)^{3/2}~,
\label{eq:jeans}
\end{equation}
where $\bar{\rho}$ is the total mass density including dark matter,
and we choose $T_{\rm IGM}=T_{\rm ad}$, the average temperature of the IGM
assuming only adiabatic cosmic expansion, consistent with our basic assumption of
not accounting for astrophysical feedback effects.


\section{Nonstandard cosmological effects and results}

\begin{figure}[htbp]
   \centering
   \includegraphics[width=16cm]{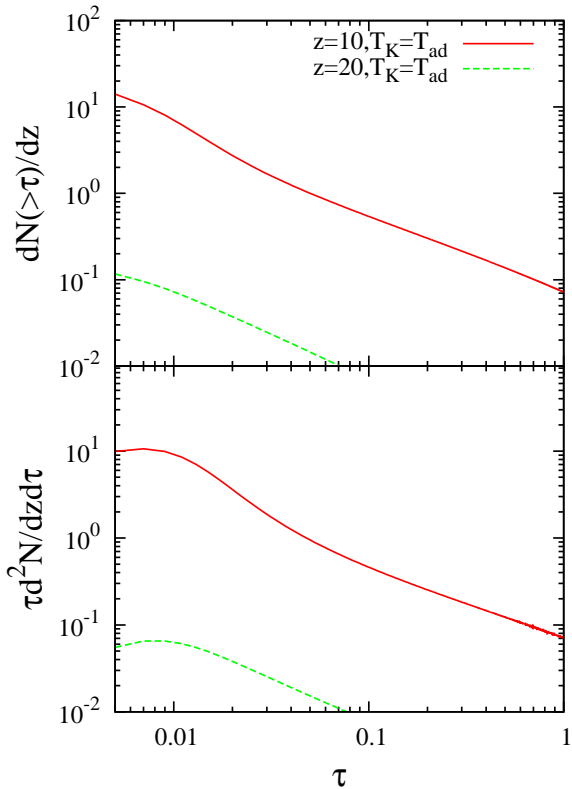} 
   \caption{{\it Top} : Cumulative abundance of 21 cm absorption features per redshift interval with optical depth greater than $\tau$
   along an average line of sight at $z=10$ (solid) and $z=20$ (dashed) for the baseline $\Lambda$CDM cosmology.
   {\it Bottom} : Same as top panel, except in terms of the differential abundance per intervals in redshift and optical depth.}
  
\label{fig:abundance}
\end{figure}

Fig.~\ref{fig:abundance} shows the abundance of 21 cm absorption features per redshift interval
along an average line of sight as a function of optical depth at $z=10$ and $20$ for the baseline $\Lambda$CDM cosmology.
Around a given $z$, the expected number of absorption features with a given $\tau$ is roughly 
$z \tau d^2N/dz d\tau$, which at $z=10$ is seen to be $\sim 100, 5, 0.7$ for $\tau \sim 0.01, 0.1, 1$, respectively,
appearing near observer frequency $\nu_{\rm obs} \sim 129$ MHz.
At $z=20$, the numbers drop considerably, simply because structure formation is less advanced compared to $z=10$,
although one can still expect $\sim 1$ absorption feature with $\tau \sim 0.01$ at $\nu_{\rm obs} \sim 68$ MHz.

\begin{figure}[htbp]
   \centering
   \includegraphics[width=16cm]{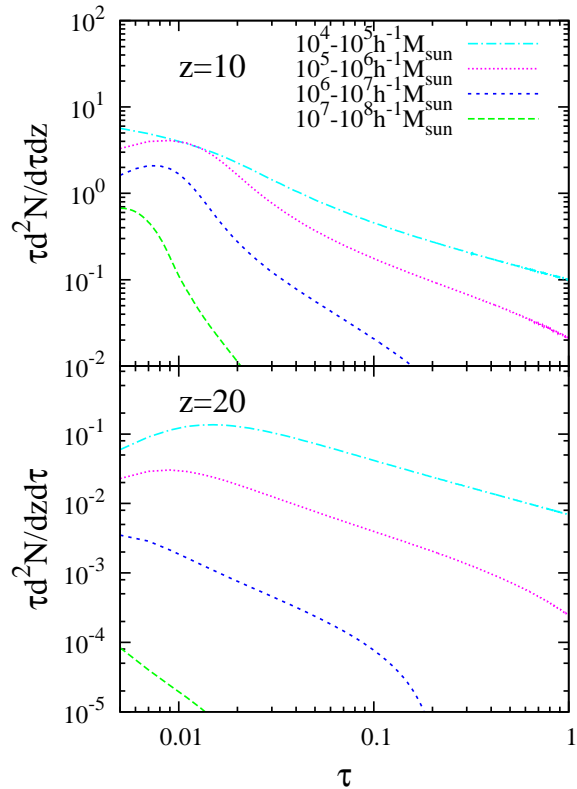}
   \caption{Contribution of halos in different mass ranges to the abundance of 21 cm absorption features
   at $z$=10 (top) and $z$=20 (bottom).
   }
\label{fig:abundance_mass}
\end{figure}

Fig.~\ref{fig:abundance_mass} compares the contributions of different ranges of halo mass 
to the absorber abundance at $z=10$ and $z=20$,
revealing that $M = 10^{4}-10^{5}h^{-1}M_{\odot}$ is most important.
With our assumption of $M_{\rm min}$ as the Jeans mass for $T_{\rm IGM}=T_{\rm ad}$,
$M_{\rm min} \sim 3.96\times 10^{4}h^{-1}M_{\odot}$ at $z$=10
and $M_{\rm min} \sim 1.68\times 10^{5}h^{-1}M_{\odot}$ at $z$=20,
so the main contribution comes from minihalos with masses just above $M_{\rm min}$.
On the other hand, minihalos with $M \sim 10^8 h^{-1}M_{\odot}$
hardly contribute to the absorbers for $\tau \gtrsim 0.01$,
on account of their higher $T_s$ as well as lower halo abundance.

In the following, we discuss how these baseline results are modified by the effects of neutrino mass, running spectral index
and warm dark matter.


\subsection{Neutrino mass}
If neutrinos have mass, the evolution of neutrino perturbations after decoupling from the hot plasma in the relativistic regime
is modified compared to the case with massless neutrinos.
The Boltzmann equations that describe the evolution of the perturbations
and the transfer function that relates initial conditions and density perturbations after recombination are changed accordingly
\cite{Lesgourgues:2006nd}. 
In addition, the energy density $\rho_{m}$ for massive neutrinos contribute as matter, as opposed to massless neutrinos that contribute as radiation. As a result, the matter power spectrum below the turnover scale
and the corresponding halo mass function are suppressed compared to the massless case.
In our calculation, we used the CAMB code for calculating the transfer function including massive neutrinos \cite{Lewis:1999bs}.

\begin{figure}[htbp]
   \centering
   \includegraphics[width=10cm]{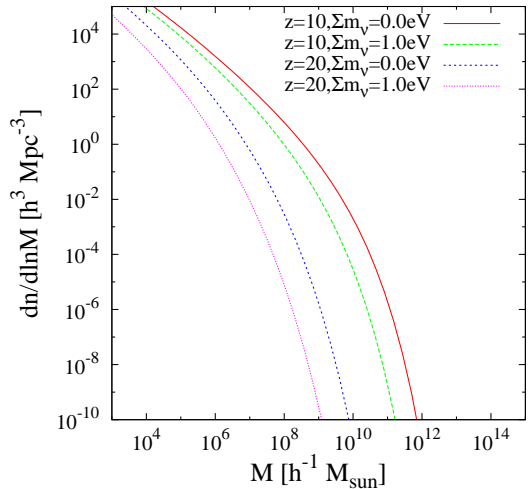}
   \caption{Halo mass functions at $z$=10 and 20 for different values of the total mass of neutrinos as indicated in the legend.}
\label{fig:mf_neutrino}
\end{figure}

The free streaming scale of neutrinos with mass $m_{\nu}$ is given by \cite{Hu:1997mj}
\begin{equation}
  k_{{\rm fs}}\sim 0.026\left(\frac{m_{\nu}}{1{\rm eV}}\right)^{1/2}\Omega_{m}^{1/2}h[{\rm Mpc}^{-1}].
\end{equation}
The suppression of the matter power spectrum below the free streaming scale is given by \cite{Hu:1997mj}
\begin{equation}
  \left(\frac{\Delta P}{P}\right)\sim -8\frac{\Omega_{\nu}}{\Omega_{m}}\sim -0.8\left(\frac{m_{\nu}}{1{\rm eV}}\right)\left(\frac{0.1N}  {\Omega_{m}h^{2}}\right).
\end{equation}
Here, $N$ is the effective number of massive neutrino species,
and $\Omega_{\nu}$ is the energy density of massive neutrinos $\rho_{\nu}$ relative to the critical density $\rho_{cr}$,
\begin{equation}
  \Omega_{\nu}=\frac{\rho_{\nu}}{\rho_{cr}}=\frac{\sum_{i}m_{i}}{93.14h^{2}[{\rm eV}]},
\end{equation}
where $m_{i}$ is the mass of each neutrino family \cite{Lesgourgues:2006nd}.
Taking all these into account, we plot the halo mass function including the effect of massive neutrinos in Fig.\ref{fig:mf_neutrino}.
It can be seen that the suppression of the mass function is most prominent at the high mass end
where it falls off exponentially.

\begin{figure}[htbp] 
   \centering
   \includegraphics[width=16cm]{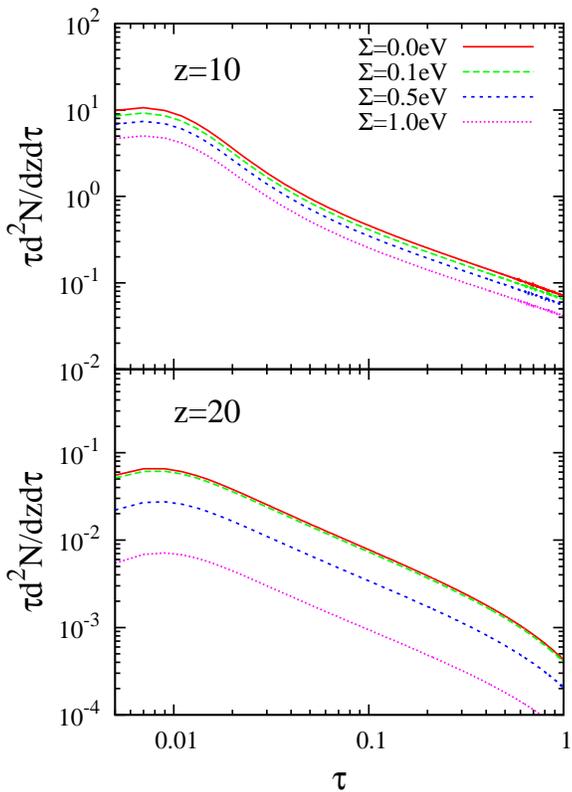} 
   \caption{Abundance of 21 cm absorption features per redshift interval at $z=10$ (top) and $z=20$ (bottom),
   for different values of the total neutrino mass $\sum m_{\nu}=0.0{\rm eV}$, $0.1{\rm eV}$, $0.5{\rm eV}$, and $1.0{\rm eV}$.}
\label{fig:abundance_neutrino}
\end{figure}

Cosmological observations have already placed relatively tight constraints on neutrino masses.
For example, upper bounds of $\sum m_{\nu} < 0.17$ eV and $\sum m_{\nu}< 0.26$ eV
have been placed from observations of the Ly$\alpha$ forest and SDSS-${\rm I}\hspace{-.1em}{\rm I}\hspace{-.1em}{\rm I}$, respectively \cite{dePutter:2012sh},
while recent {\it Planck} results give $\sum m_{\nu} < 0.66$ eV from the CMB alone \cite{Ade:2013zuv}.
To illustrate the effect of neutrino mass, we consider three
cases, $\sum m_{\nu}=0.1$, $0.5$ and $1.0$ eV.

The abundance of absorbers for different neutrino masses at $z=$10 and 20
is shown in Fig.\ref{fig:abundance_neutrino}.
At $z=10$, the resulting differences are found to be quite small, less than a factor of 3 even for $\sum m_{\nu}=1.0$ eV.
However, at $z=20$, this can become much larger,
reflecting the differences in the halo mass function in the mass range that is most important
for the 21 cm forest signal, $M \approx 10^{4} \sim 10^{6} h^{-1}M_\odot$.
Fig.\ref{fig:mf_neutrino} shows that the exponential tail of the mass function,
where the effect of the neutrino mass is largest,
is much closer to this range for $z=20$ than for $z=10$.
This is despite the fact that the suppression of the linear matter transfer function below the free streaming scale
is actually smaller at higher redshifts \cite{Ichiki:2011ue}.

At any rate, for the purpose of constraining neutrino masses,
it is apparent that the 21 cm forest must be observed at $z \sim 20$ or higher
in order to have any practical value in comparison with other methods.
We return to this issue in Section IV.


\subsection{Running spectral index}
The running of the spectral index $n_s$ of primordial fluctuations, $dn_s/d\ln k$,
is defined by
\begin{eqnarray}
  \Delta_{{\cal R}}^{2} &=& \frac{k^{3}\langle \mid{\cal R}_{k}\mid^{2} \rangle}{2\pi^{2}} \notag \\
  &=&\Delta^{2}_{{\cal R}}(k_{0})\bigg(\frac{k}{k_{0}}\bigg)^{n_{s}-1+\frac{1}{2}\ln(k/k_{0})dn_{s}/d\ln k},
\label{eq:running}
\end{eqnarray}
where $k_{0}=0.05$Mpc$^{-1}$, and ${\cal R}_{k}$ is the primordial curvature perturbation \cite{Kosowsky:1995aa}.

\begin{figure}[htbp]
   \centering
   \includegraphics[width=9cm]{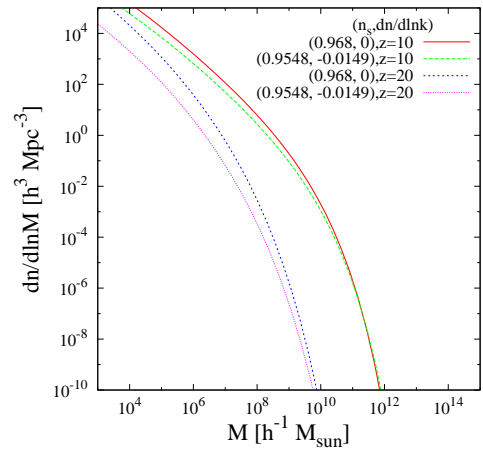} 
   \caption{Halo mass functions at $z$=10, 20 for various combinations of the spectral index $n_s$ and its running $dn_s/d\ln k$
   as indicated in the legend.}
\label{fig:mf_run}
\end{figure}

\begin{figure}[htbp]
   \centering
   \includegraphics[width=16cm]{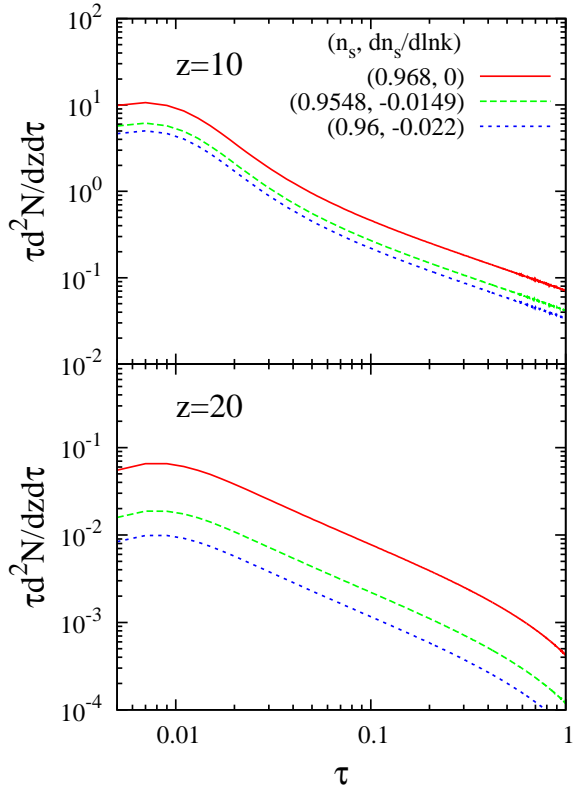} 
   \caption{Abundance of 21 cm absorption features per redshift interval at $z$=10 (top) and $z$=20 (bottom),
    for various combinations of the spectral index $n_s$ and its running $dn_s/d\ln k$ as indicated in the legend.
    Note that the case of $n_s$=0.9548 and $dn_s/d\ln k$=-0.0149 corresponds to the constraints from
    Planck + WMAP polarization + high-l CMB data.}
\label{fig:abundance_run}
\end{figure}

The latest constraints from Planck on the spectral index and RSI are
$n_s=0.9548\pm0.0073$ and $dn_s/d\ln k=-0.0149\pm 0.0085$
in combination with WMAP polarization and high-l CMB data,
and $n_s=0.9596\pm 0.0063$ and $dn_s/d\ln k=-0.0130\pm 0.0090$
in combination with WMAP polarization and BAO data \cite{Ade:2013uln}.  
For several assumed combinations of the spectral index and RSI,
we show the resulting halo mass function and abundance of 21 cm absorbers
in Fig.\ref{fig:mf_run} and Fig.\ref{fig:abundance_run}, respectively.

In contrast to massive neutrinos that suppress the power spectrum uniformly at all scales below the free streaming scale, 
the effect of RSI can be potentially more significant, since it becomes progressively larger as one goes to smaller scales.
However, taking into account the latest constraints from Planck and other observations,
we see that its effect on the 21 cm forest at $z=10$ remains within a factor or a few.
As with the case of massive neutrinos, the effect of RSI is found to be larger at $z=20$ for similar reasons:
the most relevant range of halo masses for the 21 cm forest is closer to the high mass tail of the mass function, 
which is exponentially sensitive to changes in the fluctuation amplitude caused by RSI.
Thus, if the 21 cm forest is observable at $z=20$ or higher, one may hope to obtain valuable constraints on RSI,
which in turn may help in discriminating different inflation models \cite{Adshead:2010mc}, 
independently from other observations.


\subsection{Warm dark matter}
To evaluate the halo mass function in the WDM cosmology,
we utilize the prescription of Smith $\&$ Markovic \cite{Smith:2011ev}.
For WDM of particle mass $m_{{\rm WDM}}$ and density $\Omega_{{\rm WDM}}$ relative to the critical density,
the comoving free streaming scale can be approximated by
\begin{equation}
  \lambda_{{\rm fs}}\sim 0.11\bigg(\frac{\Omega_{{\rm WDM}}h^{2}}{0.15}\bigg)^{1/3}
  \bigg(\frac{m_{{\rm WDM}}}{{\rm  keV}}\bigg)^{-4/3} [{\rm Mpc}].
\label{eq:free_streaming}
\end{equation}
The mass scale below which halo formation is suppressed is \cite{AvilaReese:2000hg}
\begin{equation}
  M_{{\rm fs}}=\frac{4}{3}\pi \bigg(\frac{\lambda_{{\rm fs}}}{2}\bigg)^{3}\bar{\rho}_{m}.
\label{eq:free_streaming2}
\end{equation}
The halo mass function in the WDM cosmology is approximately \cite{Smith:2011ev}
\begin{equation}
  \frac{dn}{dM}(M,z)=\frac{1}{2} \bigg \{1+{\rm erf}
  \bigg[\frac{\log_{10}(M/M_{{\rm fs}})}{\sigma_{\log M}}\bigg] \bigg \}
  \bigg[\frac{dn}{dM} \bigg]_{{\rm PS}}~.
\label{eq:wdm_mass}
\end{equation}
Here $\sigma_{\log M}$=0.5,
and $[dn/dM]_{{\rm PS}}$ is the Press-Schechter mass function evaluated
with a fitting formula for the matter power spectrum with WDM \cite{Bode:2000gq,Viel:2005qj}
\begin{equation}
  P_{{\rm WDM}}(k)=P_{{\rm CDM}}(k)\{[1+(\alpha k)^{2\mu}]^{-5/\mu}\}^{2}~,
\end{equation}
where $\alpha$ and $\mu$ are fitting parameters given by
\begin{equation}
  \alpha=0.049\bigg(\frac{m_{{\rm WDM}}}{{\rm keV}}\bigg)^{-1.11}\bigg(\frac{\Omega_{{\rm WDM}}}{0.25}\bigg)^{0.15}
  \bigg(\frac{h}{0.7}\bigg)^{1.22}h^{-1} [{\rm Mpc}]~
\label{eq:alpha}
\end{equation}
and $\mu=1.12$ \cite{Viel:2005qj}.

\begin{figure}[htbp]
   \centering
   \includegraphics[width=16cm]{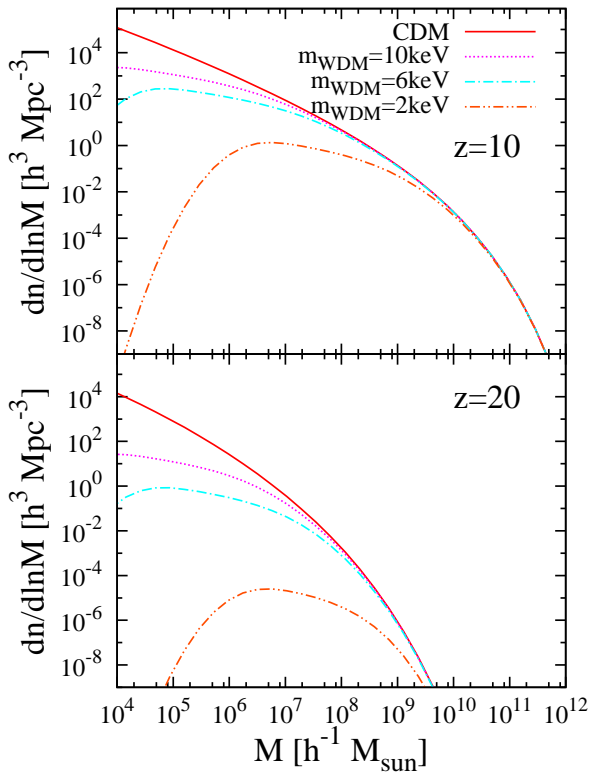} 
   \caption{Halo mass functions at $z=10$ (top) and $z=20$ (bottom)
   for CDM (red), and WDM with $m_{{\rm WDM}}=10$ keV (blue) and $2$ keV (green).}
\label{fig:mf_wdm}
\end{figure}

The resulting halo mass functions at $z=10$ and $20$
for WDM with different particle masses compared with CDM
are plotted in Fig.\ref{fig:mf_wdm}.
As can clearly be seen,
WDM drastically suppresses the mass function below the mass scale $M_{{\rm fs}}$ that depends on $m_{\rm WDM}$,
while remaining identical to CDM above this scale.

\begin{figure}[htbp]
   \centering
   \includegraphics[width=16cm]{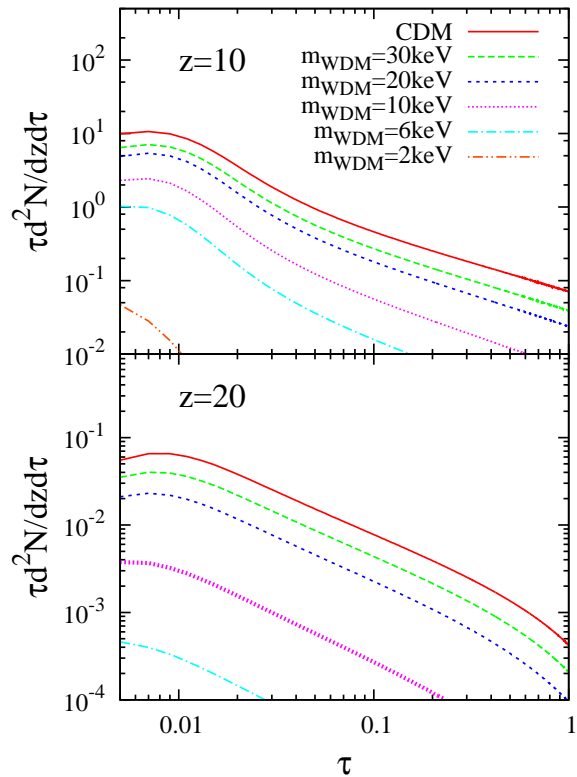} 
   \caption{Abundance of 21 cm absorption features per redshift interval at $z=10$ (top) and $z=20$ (bottom) 
   for WDM with various particle masses as indicated in the legend.}
\label{fig:abundance_wdm}
\end{figure}

Fig.\ref{fig:abundance_wdm} shows the corresponding abundance of 21 cm absorbers for WDM,
which manifest dramatic changes in accord with the halo mass function at small masses.
The effects at $z=20$ are even stronger than at $z=10$, for reasons similar to that discussed above for neutrinos or RSI.
In fact, if $m_{\rm WDM}$ is in the few keV range as is favored to explain the missing satellite problem \cite{Polisensky:2010rw},
the suppression would be so great as to virtually make any 21 cm forest signal unobservable,
as the relevant minihalos are much smaller than the satellites in question.

On the other hand, the cause of the missing satellite problem may lie in some kind of astrophysical feedback effect.
From a particle physics perspective, WDM is still motivated in some theories, for example as sterile neutrinos,
whose mass has been constrained to be in the range $\sim$ 1-50 keV, also depending on the mixing angle
(see Fig.2 in \cite{Boyarsky:2012rt}).
The current upper limits on their mass come from non-detections of X-ray lines caused by decaying sterile neutrinos
in clusters of galaxies and the cosmic X-ray background.
We see from Fig.\ref{fig:abundance_wdm} that future observations of the 21 cm forest may provide
an observable and potentially more sensitive probe of such particles with masses $m_{\rm WDM}\gtrsim 10$ keV.

We note recent observational evidence of weak, unidentified X-ray lines from some clusters and galaxies,
announced after submission of this paper \cite{Boyarsky:2014jta}. 
One possible interpretation is in terms of decaying sterile neutrinos with mass $\sim 7{\rm keV}$,
which, if confirmed, may be challenging to probe with the method proposed here.
However, other interpretations are possible and these observations cannot yet be taken
as definitive evidence of WDM or measurement of its mass.


\section{Discussion and Summary}

We now turn to a discussion of the observability of the 21~cm forest due to minihalos.
The principal question is the existence of background radio sources
with sufficient brightness and number at the relevant frequency and redshifts of $z \sim 10-20$.
The low temperatures of minihalos imply that the width of the expected absorption features
are narrow, necessitating spectroscopy with frequency resolution of order $\Delta \nu \sim$ kHz
at observer frequencies $\nu_{\rm obs} \sim$ 70-130 MHz.
Following and updating \cite{Furlanetto:2002ng},
in order to detect absorption features of optical depth $\tau$ with frequency resolution $\Delta \nu$
and signal-to-noise S/N with an integration time $t_{\rm int}$,
the required minimum background source brightness is

\begin{equation}
\begin{split}
  S_{\rm min}&=10.4\rm{mJy}\bigg(\frac{0.01}{\tau}\bigg)\bigg(\frac{S/N}{5}\bigg)\bigg(\frac{\rm{1 kHz}}{\Delta \nu}\bigg)^{1/2}\\
  &\times\bigg(\frac{5000[\rm m^{2}/K]}{A_{\rm{eff}}/T_{\rm{sys}}}\bigg)\bigg(\frac{100~\rm{hr}}{t_{{\rm int}}}\bigg)^{1/2},
\end{split}
\end{equation}
where the specifications anticipated for SKA2-low are adopted for
the effective collecting area $A_{{\rm eff}}$ and system temperature $T_{{\rm sys}}$
\cite{Huynh:2013aea, Ciardi2014SKAbook}.

Our results in Section III at face value show that spectroscopy of a single source with such properties at $z \sim 10$
may reveal tens to hundreds of absorption features with $\tau \sim 0.01-0.1$,
which could already provide important information on the SSPS.
Multiple sources would still be desirable to characterize fluctuations along different lines of sight.
On the other hand, at $z \sim 10$, our neglect of astrophysical effects
such as the UV background or reionization and heating of the IGM is hardly justifiable.
As mentioned below, in reality, such effects may completely dominate over any of the
SSPS-related effects discussed above,
which were quite small already at $z=10$ except for the case of WDM.

In this regard, $z \sim 20$ or higher would be much more preferable, since the formation of stars and galaxies
and their consequent feedback effects are likely to be considerably less advanced.
Moreover, as seen in the previous section, the effects on the 21 cm forest caused by interesting non-standard physics aspects
such as neutrino mass, running spectral index and warm dark matter all become significantly larger at $z \sim 20$.
On the other hand, the expected number of absorption features is much less,
only of order one with $\tau \sim 0.01$ along a given line of sight.
Thus, at these redshifts, at least several (and preferably much more) background sources would be required
for the 21 cm forest to be a useful probe of the SSPS.

Provided that such measurements can be successfully conducted,
the uncertainties with which the key physical parameters are constrained
may be roughly estimated as follows.
Focusing on $z=20$, the bottom panels of Figs. 6, 8 and 10 show that the most essential observable for our objectives
is the abundance of absorption features with $\tau \sim 0.01$.
From the bottom panels of Figs. 2 and 4, it can be judged that such absorbers reflect the most probable lines of sight
through minihalos in the crucial mass range just above $M_{\rm min}$ (Section III).
Since we foresee on average only one such absorber per line of sight out to $z=20$ for our baseline case, 
the main uncertainty would come from the number of available background radio sources with flux $S > S_{\rm min}$.
Assuming Poisson statistics, measurements for $N$ lines of sight would imply uncertainties of $1/\sqrt{N}$;
for example, if suitable observations can be made for 100 background sources at $z=20$,
the parameters characterizing the SSPS can be constrained to an accuracy of 10\%.

Primary candidates for such sources at high redshifts are radio-loud quasars.
For example, an object similar to a powerful, local radio galaxy such as Cyg A
would have the requisite brightness if placed at $z \sim 10$ \cite{Carilli:2002ky}.
Estimates based on extrapolations of the observed radio luminosity functions to higher redshifts
suggest that depending on the assumptions,
there could be as many as $\sim 10^4-10^5$ and $\sim 10^3-10^4$ radio quasars
with sufficient brightness in the whole sky at $z=10$ and $z=15$, respectively \cite{Xu:2009dr}
(see also \cite{Haiman:2004ny}).
However, from a physical standpoint, it is an open question whether black holes with accordingly large masses
could already have existed at such epochs.

An alternative possibility is the radio afterglows of certain types of GRBs.
GRBs have already been observed up to $z \sim 8-9$, and it is plausible that they occur up to the earliest epochs
of star formation in the universe at $z \sim 20$ or higher \cite{Bromm:2007dq}.
However, if such GRBs are similar to those seen at lower redshifts, their radio afterglows
are not expected to be bright enough at the relevant observer frequencies $\nu_{\rm obs} \sim$ 100 MHz
due to strong synchrotron self-absorption \cite{Inoue:2003ga}. 
On the other hand, it has been recently proposed that GRBs arising from Population (Pop) III stars
forming in metal-free environments may be much more energetic compared to ordinary GRBs,
which can generate much brighter low-frequency radio afterglows by virtue of their blastwaves expanding to larger radii
over longer timescales $t_{\rm rad,pk} \sim 1000$ yr \cite{Toma:2010xc}.
If the rate of Pop III GRBs with sufficiently bright radio emission is 0.1 yr$^{-1}$ or roughly $10^{-4}$ of all GRBs,
one can expect $\sim 100$ such sources all sky at a given time.
Thus they may potentially suffice for 21 cm forest studies even at $z \sim 20$, albeit with large uncertainties.
A practical question that remains is how we can observationally identify such sources.
Further discussions on the observability of the 21 cm forest are beyond the scope of this paper
and will be explored in future work.

\begin{figure}[htbp]
  \centering
   \includegraphics[width=8cm]{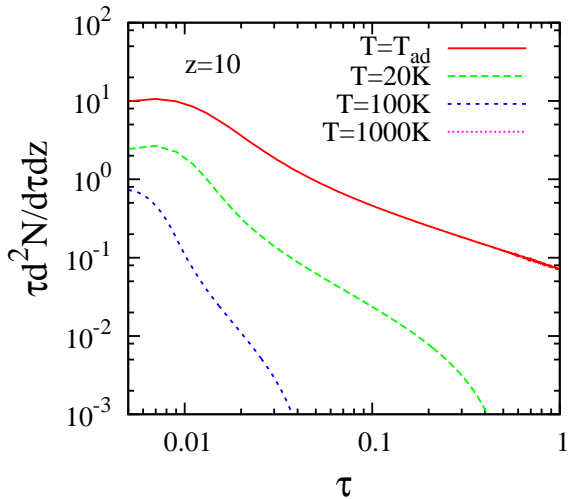}
   \caption{Abundance of 21 cm absorption features per redshift interval at $z$=10 for different values of $T_{\rm IGM}$ as indicated in the legend.}
\label{fig:abundance_temp}
\end{figure}

Next, we briefly discuss some aspects of astrophysical feedback effects
that we have chosen to neglect in this work in order to focus on the implications of the SSPS.
Once the formation of stars and/or black holes is initiated in the universe,
a background of UV and X-ray photons will build up over time.
Ly $\alpha$ photons can resonantly scatter with hydrogen atoms and alter its hyperfine excitation state
via the Wouthuysen-Field effect (see Eq.\ref{eq:spin1}) \cite{Furlanetto:2006jb}.
Furthermore, UV and X-ray photons as well as shocks driven by supernova explosions, quasar outflows, etc.
can heat the IGM to temperatures much above our assumed value of $T_{\rm ad}$
corresponding to simple, adiabatic cosmic expansion.
The consequences of such effects on the 21 cm forest are likely to be significant
\cite{Furlanetto:2002ng, Carilli:2002ky, Furlanetto:2006dt, Xu:2009dr,
Xu:2010us, Meiksin:2011gx, Mack:2011if, Ciardi:2012ik, Ewall-Wice:2013yta},
especially at $z \sim 10$, where it is clear that cosmic reionization
is already in progress from CMB polarization measurements.
As a simple illustration of such feedback effects, Fig.\ref{fig:abundance_temp} shows
how the 21 cm forest at $z$=10 is affected by introducing a uniform temperature floor in the IGM at different values.
The main consequence here is the increase of the Jeans mass, which eliminates the smaller minihalos
that are predominantly responsible for the 21 cm forest signal and leads to its severe suppression.
Compared to the effects of the SSPS discussed in Section III, those due to feedback exhibit a much stronger
dependence on $\tau$, which in principle may help in distinguishing the two.
However, in practice, exploring the SSPS clearly favors observations at $z \sim 20$ and above
where such feedback effects are expected to be more limited,
in addition to the fact that the SSPS-related effects are larger, including those
caused by massive neutrinos, running spectral index and warm dark matter.

We comment on a few other pertinent issues that have not been fully addressed in this paper.
First, at least for low redshifts, the Press-Schechter form of the halo mass function that we have adopted
is not known to be the most precise representation of numerical N-body simulation results,
and other forms such as that of Sheth and Tormen \cite{Sheth:1999mn} are more often employed in the literature.
However, the situation is currently less clear for the high redshifts of our interest \cite{Reed:2006rw, Lukic:2007fc, Iliev:2010cy},
so the Press-Schechter form was chosen here for simplicity.
Further progress warrants more conclusive studies on the subject.
Second, our treatment of the cosmological effects induced by massive neutrinos, RSI or WDM
concentrated on the resultant modifications to the halo mass function.
On the other hand, these effects can also alter the dark matter profile within individual halos
and hence the gas profile as well, potentially affecting the 21 cm absorption signal.
Consequences for the halo dark matter profile of massive neutrinos \cite{Brandbyge:2010ge},
RSI \cite{Fedeli:2010iw} and WDM \cite{Schneider:2011yu, Markovic:2013iza}
have been studied to some extent for galaxy- to cluster-scale halos at low redshifts,
indicating that they are affected mainly in the central regions with $r/r_{{\rm vir}} \lesssim$ 0.1.
Although no corresponding work exists for high-redshift minihalos,
we may speculate that the impact is less than that due to the mass function
with regard to our results, for which the outer regions of the halo are more relevant (Fig.2).
Nevertheless, this needs to be substantiated by future, dedicated investigations.
Third, we did not account for neutral gas lying outside the virial radii of minihalos and accreting onto them,
which can provide a significant additional contribution to the absorption feature \cite{Furlanetto:2002ng, Xu:2010us}.
Albeit challenging to model accurately, such components should be taken into account for more accurate predictions in the future.
Note also the possibility of further absorption along the line of sight
due to the incompletely virialized cosmic web and/or the global IGM that is expected to be much weaker
\cite{Carilli:2002ky, Furlanetto:2006dt, Mack:2011if, Ciardi:2012ik},
and that due to the disks of larger galaxies that should be individually stronger but much rarer \cite{Furlanetto:2002ng}.
Finally, the implications of relative streaming velocity between baryons and dark matter
\cite{Tseliakhovich:2010yw} 
may also be interesting for future studies of the 21 cm forest.

To conclude,
we have presented a novel approach to probe small-scale cosmological fluctuations utilizing the 21 cm forest,
that is, absorption features caused by H{\sc I} gas in minihalos
in the spectrum of background radio sources at redshifts at $z \sim 10$ and above.
The method is potentially sensitive to scales $k\gtrsim 10$ Mpc$^{-1}$,
much smaller than can be currently studied via observations of the CMB, galaxy clustering or the Ly$\alpha$ forest.
New insight can be expected into aspects of physics beyond the standard $\Lambda$CDM cosmological model
such as massive neutrinos, running of the primordial spectral index and warm dark matter.
Radio quasars or Population III gamma-ray bursts are potential candidates for the background radio sources
with the requisite brightness and number at the appropriate redshifts for future observations with SKA.

Further potentially interesting cosmological applications of the 21 cm forest
include probes of primordial non-Gaussianity in relation to either
the nonlinear, scale-dependent bias  \cite{Chongchitnan:2012we}
or the halo mass function \cite{Matarrese:2000iz}, 
and probes of isocurvature primordial perturbations (e.g \cite{Linde:1996gt}). 
We note that several recent papers have discussed the possibility of studying various aspects of the SSPS
via the 21 cm emission signal \cite{Oyama:2012tq},
although efficient removal of the far brighter foreground emission poses a major observational challenge
for realizing such prospects \cite{Oh:2003jy}.

\section*{ACKNOWLEDGMENT}
We would like to thank Keitaro Takahashi, Tsutomu T. Takeuchi,
Kyungjin Ahn, Matt Jarvis, Takamitsu Tanaka, Eiichiro Komatsu,
Yidong Xu and Benedetta Ciardi
for helpful comments and useful discussions.
This work has been supported in part
by Grant-in-Aid for Scientific Research Nos. 25-3015(HS), 24340048 (KI \& SI), 24-2775 (SY)
from the Ministry of Education, Sports, Science and Technology (MEXT) of Japan,
and by Grant-in-Aid for the Global Center of Excellence program at Nagoya University
"Quest for Fundamental Principles in the Universe: from Particles to the Solar System and the Cosmos"
from the MEXT of Japan.

\end{document}